\PassOptionsToPackage{table}{xcolor}
\documentclass{article}

\usepackage{spconf, amsmath, graphicx}

\usepackage{appendix, color, hyperref, multicol, soul, svg}
\usepackage[hang,flushmargin]{footmisc}

\title{Neural Pitch-Shifting and Time-Stretching with Controllable LPCNet}
\name{Max Morrison$^{\flat}$, Zeyu Jin$^{\natural}$, Nicholas J. Bryan$^{\natural}$, Juan-Pablo Caceres$^{\natural}$, Bryan Pardo$^{\flat}$\thanks{This material is based upon work supported by the National Science Foundation Graduate Research Fellowship under Grant No. DGE-1842165. \newline
This work has made use of the Mystic (Programmable Systems Research Testbed to Explore a Stack-WIde Adaptive System fabriC) NSF-funded infrastructure at Illinois Institute of Technology, NSF award CRI-1730689.}} 
\address{$^{\flat}$Northwestern University, Evanston, IL, USA\\
         $^{\natural}$Adobe Research, San Francisco, CA, USA}

\begin{document}
\ninept
\maketitle


\begin{abstract}
Modifying the pitch and timing of an audio signal are fundamental audio editing operations with applications in speech manipulation, audio-visual synchronization, and singing voice editing and synthesis. Thus far, methods for pitch-shifting and time-stretching that use digital signal processing (DSP) have been favored over deep learning approaches due to their speed and relatively higher quality. However, even existing DSP-based methods for pitch-shifting and time-stretching induce artifacts that degrade audio quality. In this paper, we propose Controllable LPCNet (CLPCNet), an improved LPCNet vocoder capable of pitch-shifting and time-stretching of speech. For objective evaluation, we show that CLPCNet performs pitch-shifting of speech on unseen datasets with high accuracy relative to prior neural methods. For subjective evaluation, we demonstrate that the quality and naturalness of pitch-shifting and time-stretching with CLPCNet on unseen datasets meets or exceeds competitive neural- or DSP-based approaches.
\end{abstract}




\section{Introduction}
\label{sec:intro}

Speech manipulation algorithms that modify the fundamental frequency and duration of speech are essential for a variety of speech editing applications, such as audio-visual synchronization, prosody editing, auto-tuning, and voice conversion. General-purpose audio editing software such as Pro Tools and Adobe Audition contains algorithms for pitch-shifting. However, these algorithms can alter the timbre of speech to sound unnatural. This motivates the development of natural-sounding pitch-shifting algorithms catered to speech.

Related speech manipulation algorithms include both digital signal processing (DSP) approaches as well as neural networks. DSP-based approaches include TD-PSOLA~\cite{Moulines_1990}, WORLD~\cite{Morise_2016}, and STRAIGHT~\cite{Banno_2007}. These methods benefit from fast inference and accurate control, but often degrade the signal with noticeable artifacts. Prior studies~\cite{morrison2020controllable} have shown TD-PSOLA to be preferable over WORLD for speech manipulation, and WORLD has been shown to be significantly preferable over STRAIGHT for speech resynthesis~\cite{morise2018sound}.

Prior methods in neural pitch-shifting include Pitch-Shifting WaveNet (PS-WaveNet)~\cite{morrison2020controllable}, Quasi-Periodic Parallel WaveGAN (QP-PWG)~\cite{Wu_2021}, Unified Source-Filter GAN (uSFGAN)~\cite{yoneyama2021unified}, and Hider-Finder-Combiner (HFC)~\cite{webber2020hider}. PS-WaveNet is too computationally expensive for our use case of real-time interactive editing. QP-PWG and uSFGAN exhibits constant-ratio pitch-shifting quality on par with WORLD, but with worse accuracy. HFC demonstrates variable-ratio pitch-shifting performance with worse accuracy than WORLD, and its subjective quality is significantly degraded by noise induced during vocoding. Only QP-PWG and uSFGAN support multiple speakers, but do not demonstrate an ability to perform variable-rate pitch-shifting. Further, none of these works propose methods for time-stretching. As well, our results indicate that WORLD can perform substantially more accurate pitch shifting than previous studies have reported~\cite{morrison2020controllable, Wu_2021, yoneyama2021unified, webber2020hider}. We hypothesize that prior methods performed improper interpolation of WORLD parameters or evaluated pitch error in unvoiced regions. 

\begin{figure}[t]
    \centering
    \includegraphics[width=0.96\linewidth]{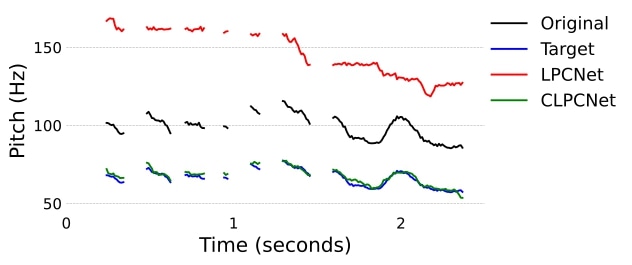}
    \caption{Pitch-shifting with LPCNet and our proposed CLPCNet. The target pitch is two-thirds the original pitch.}
    \label{fig:overview}
    \vspace{-6mm}
\end{figure}

 Neural vocoders are deep neural networks that convert acoustic features (e.g., a mel-spectrogram) to a waveform. Using a neural vocoder, we can perform speech manipulation by encoding speech audio as acoustic features, modifying these acoustic features, and then vocoding to produce a new waveform. Recent neural vocoders include WaveGlow~\cite{prenger2019waveglow}, Parallel WaveGAN~\cite{yamamoto2020parallel}, Neural Source Filter (NSF)~\cite{wang2019neural}, and LPCNet~\cite{valin2019lpcnet}. None of these methods address pitch-shifting or time-stretching except LPCNet, which has been informally shown to be able to perform time-stretching~\cite{valin2019blog}. However, no evaluation of time-stretching performance is provided. LPCNet resembles a source-filter model, which decouples the residual (pitch and noise) and spectral (timbre) structure. Source-filter models are usually capable of pitch-shifting speech, exhibiting more natural timbre than traditional phase vocoders~\cite{Quatieri2001}.  However, our work demonstrates that, without modification, LPCNet does not perform accurate pitch-shifting (Figure~\ref{fig:overview}). We hypothesize this is due to three issues: (1) limitations in the pitch representation used in LPCNet, (2) insufficient disentanglement between pitch and acoustic features, and (3) a lack of training data for very high- and low-pitched speech. Kons et al.~\cite{kons2019high} sidestep these limitations by generating the input parameters using a separate neural network. However, their approach necessitates training multiple neural networks and does not generalize to unseen speakers without speaker adaptation.

Rather than sidestepping these limitations, as Kons et al. do, we directly address them to create Controllable LPCNet (CLPCNet), which significantly improves the synthesis quality and pitch-shifting performance of LPCNet. In our objective evaluation, we show that CLPCNet performs both constant- and variable-ratio pitch-shifting and time-stretching with high accuracy on unseen speakers and datasets. In our subjective evaluation, we show that the quality of pitch-shifting and time-stretching with CLPCNet meets or exceeds competitive DSP-based methods. CLPCNet also substantially improves the quality of speech vocoding compared to LPCNet, and permits simultaneous speech coding and speech manipuation. Code is available under an open-source license at \url{https://github.com/maxrmorrison/clpcnet}.\footnote{Audio examples are available at \url{https://main.d3ee4zjxcj59ad.amplifyapp.com/}.}


\section{LPCNet}

LPCNet~\cite{valin2019lpcnet} is a neural vocoder that models each sample of a speech signal as the sum of a deterministic term (the \textit{prediction}) and a stochastic term (the \textit{excitation}). The prediction is computed via linear predictive coding (LPC)~\cite{makhoul1975linear}, where LPC coefficients are derived from Bark-frequency cepstral coefficients (BFCCs). LPCNet autoregressively predicts the parameters of a categorical distribution over 8-bit mu-law-encoded excitation values.

\subsection{Architecture}
\label{sec:lpcarc}

LPCNet consists of two subnetworks: the frame-rate and sample-rate networks. The frame-rate network consists of a pitch embedding layer followed by two 1D convolution layers with tanh activations and two dense layers with tanh activations. The sample-rate network consists of an embedding layer for sample-rate features followed by two gated recurrent units (GRUs) with sigmoid and softmax activations, respectively. The frame-rate network takes as input the YIN~\cite{de2002yin} pitch, pitch correlation~\cite{valin2018hybrid} (henceforth referred to as \textit{periodicity}), and 18-dimensional BFCCs with a hop size of 10 milliseconds and produces a 128-dimensional embedding for each frame. The sample-rate network takes four inputs:  (1) the previously generated excitation, (2) the previous sample value, (3) the current prediction value (see previous paragraph), and (4) the output of the last layer of the frame-rate network after nearest neighbors upsampling.

\subsection{Time-stretching}

The time resolutions of the sample-rate and frame-rate networks are related by upsampling factor $k$; for every frame processed by the frame-rate network, the sample-rate network produces $k$ samples without overlap between frames. LPCNet can perform time-stretching by using a variable-rate hop size $k_f$ on a per-frame basis. For example, if a phoneme is spoken for 100 milliseconds (10 frames), we can stretch the phoneme to 200 milliseconds by decoding twice as many samples from each frame.


\section{Controllable LPCNet}

While LPCNet achieves competitive audio quality and time-stretching performance, it is unable to perform accurate pitch-shifting (see Figure~\ref{fig:overview}). Below, we elaborate on our hypothesis of the three issues prohibiting pitch-shifting in LPCNet (see Section~\ref{sec:intro}) and propose solutions to these issues. In addition, we propose a simplification of the sampling procedure of LPCNet (see Section~\ref{sec:sample}).

\subsection{Pitch representation}
\label{sec:pitch}


We identify two issues with the pitch representation used in LPCNet. First, pitch values are encoded as the number of samples per period. This design makes pitch bins perceptually uneven; higher frequencies are coarsely sampled, with some bin widths exceeding 50 cents. Given 8-bit quantization at a sample rate of 16 kHz, the minimum representable frequency is 63 Hz, which prohibits modeling very low-pitched voices. We propose a quantization of the frequency range 50-550 Hz that is equally spaced in base-2 log-scale, which makes the width of each bin 16.3 cents. 


Second, the YIN pitch and periodicity exhibit significant noise, which harms the performance of LPCNet. Therefore, we use CREPE~\cite{kim2018crepe} (specifically \texttt{torchcrepe}~\cite{morrison2020torchcrepe}) to extract the pitch and periodicity. CREPE outputs a distribution over quantized pitch values over time. We apply Viterbi decoding~\cite{forney1973viterbi} to extract a smooth pitch trajectory, which reduces half and double frequency errors. We dither the extracted pitch with random noise drawn from a triangular distribution centered at zero, with width equal to two CREPE pitch bins (i.e., 40 cents). This reduces quantization error without increasing the noise floor~\cite{vanderkooy1987dither}. Our CREPE periodicity measure is the sequence of probabilities associated with the pitch bins selected by Viterbi decoding. CREPE normalizes each frame of input audio, making it invariant to amplitude. This causes low-bit noise to be labeled as periodic during silent regions. We avoid this by setting the periodicity to zero in frames where the A-weighted loudness~\cite{mccurdy1936tentative} is less than -60 dB, relative to a reference of 20 dB. Our periodicity measure has a correlation of .82 with the periodicity measure of YIN, and visual inspection indicates that our representation contains significantly less noise (see companion website). This indicates that CREPE learns a representation of speech periodicity at least as good as autocorrelation-based methods.

\subsection{Data augmentation}
\label{sec:augment}

We perform training with a much larger dataset than the original LPCNet (see Section~\ref{sec:data}). For this reason, we omit the original data augmentation, which includes random biquad filtering, volume augmentation, and noise injection. Instead, we propose a novel augmentation to improve pitch-shifting performance.

High-accuracy pitch-shifting with a neural network requires that the input pitch representation is disentangled from other features (e.g., the BFCCs). As well, values close to 50 or 550 Hz are rarely found within speech datasets, which prohibits the network from learning to pitch-shift to these values. We propose a resampling data augmentation to better disentangle pitch features from the BFCCs and allow pitch-shifting of speech to pitch values not seen in the training data. Let $\phi(*; a, b)$ be a function that resamples a signal from sampling rate $a$ to sampling rate $b$. Given speech signal $x$ with original sampling rate $s$, target sampling rate $t = 16000$, and constant pitch shift factor $r$, we augment training data with $x_r = \phi(\phi(x; rs, s); s, t)$ for values of $r$ in $[ \frac{1}{2}, \frac{2}{3}, \frac{3}{4}, \frac{4}{5}, \frac{5}{4}, \frac{4}{3}, \frac{3}{2}, 2 ]$. Performing pitch-shifting at the original sampling rate ensures that we do not lose high-frequency information when downsampling. This resampling method significantly modifies the speech formants. We hypothesize that this encourages the model to disentangle pitch from the representation of formants within the BFCCs.

\subsection{Sampling excitation values}
\label{sec:sample}

The original LPCNet samples excitation values with sampling temperature dependent on the periodicity. We instead use a constant sampling temperature of 1, which we find performs equivalently when the amount of training data is sufficiently large. We retain the thresholding of the distribution at small values. Let  $p(e_t = c)$ for $c=1, \dots, 256$ be the predicted 256-dimensional categorical distribution over mu-law-encoded excitation values. Let $P_{t, c} = \max[0, p(e_t = c) - T]$, where $T$ is a constant threshold. We sample excitations from the categorical distribution $P_{t, c} / \sum_{i=1}^{256} P_{t,i}$. We use $T=.001$, which maximizes the F1 score of the voiced/unvoiced decision.


\section{Evaluation}

We design our evaluation to test two hypotheses: (1) CLPCNet allows users to perform pitch-shifting and time-stretching of speech with high accuracy suitable for prosody modification, and (2) the subjective quality of pitch-shifting and time-stretching with CLPCNet meets or exceeds that of TD-PSOLA~\cite{Moulines_1990} and WORLD~\cite{Morise_2016}, two competitive DSP-based methods that have not been outperformed by existing neural methods. We use the Python \texttt{psola}~\cite{morrison2021psola} and \texttt{pyworld}~\cite{hsu2021pyworld} packages as baseline for TD-PSOLA and WORLD, respectively. We use CREPE to extract pitch contours used to control both CLPCNet, TD-PSOLA, and WORLD.  We also compare to the original LPCNet model (using checkpoint \texttt{lpcnet20h\_384\_10\_G16\_80.h5} in the public LPCNet implementation~\cite{valin2019github}). In our objective evaluation, we ablate each of our three proposed improvements to LPCNet, corresponding to sections \ref{sec:pitch} through \ref{sec:sample}. \textbf{For all tables, $\uparrow$ means higher is better and $\downarrow$ means lower is better.}

\subsection{Data}
\label{sec:data}

We use the VCTK dataset~\cite{yamagishi2019cstr} for training. We train on 100 speakers, withholding four male and four female speakers for unseen speaker evaluation. To evaluate on unseen utterances by speakers seen during training, we set aside four utterances per speaker from four female and four male speakers in the training data. We use \texttt{microphone 2}, which contains less distortion and noise. To test the robustness of CLPCNet to unseen recording conditions, we perform additional evaluation on the clean partition of the DAPS dataset~\cite{mysore2014can} as well as the RAVDESS~\cite{livingstone2018ryerson} dataset. RAVDESS contains a significant amount of reverb, which we remove using HiFi-GAN~\cite{su2020hifi}.

We resample all audio to 16 kHz and apply a 5th-order Butterworth high-pass filter with a 65 Hz cutoff to remove the 50 Hz hum in VCTK. This filter is shallow enough for CLPCNet to perform accurate pitch-shifting below the cutoff (e.g., see Figure~\ref{fig:overview}).  We apply a preemphasis filter with a coefficient of .85, followed by a limiter to prevent clipping~\cite{bechtold2015limiter}. CREPE pitch is extracted from the audio prior to preemphasis. As in the original LPCNet, YIN pitch is extracted after preemphasis. We found that peak normalization to 0.8 or 1.0 as well as LUFS normalization~\cite{lufs} all harmed performance, but without normalization, examples with low peak amplitude have artifacts in voiced regions. Therefore, we normalize utterances with a peak amplitude less than 0.2 to have a peak amplitude of 0.4.

\subsection{Training}

We train CLPCNet for 45 million steps with a batch size of 64. The number of steps was selected to maximize the F1 score of voiced frame classification. Each item in the batch contains a random slice of 15 frames of BFCCs and pitch features and the corresponding 2400 excitation, prediction, and sample features. We use the AMSGrad~\cite{reddi2019convergence} optimizer with a learning rate of $10^{-3}$ and weight decay of $5 \times 10^{-5}$ to minimize the cross entropy loss between the predicted and ground truth excitations. We omit sparsifying the GRU weights, which does not harm quality when the dataset is sufficiently large.

\subsection{Objective evaluation}
\label{sec:objec}

We report objective metrics to measure the ability of CLPCNet to perform constant- and variable-ratio pitch-shifting. We omit objective evaluation of time-stretching, as the generated audio is precisely sample-aligned by construction. Both LPCNet and CLPCNet take as input a pitch contour (see Section~\ref{sec:lpcarc}). To measure pitch accuracy, we replace the input pitch with a target pitch and compare the pitch of the synthesized audio to the target pitch. We report three objective pitch metrics: (1) \textbf{RMS}, the root-mean-square of the pitch error in cents within frames where both the target and synthesized speech are classified as voiced, (2) \textbf{F1}, the F1 score of the binary voiced/unvoiced decision, and (3) \textbf{GPE}, the gross pitch error, defined as the fraction of voiced pitch values with pitch error greater than $k$ cents. We use $k = 50$. For constant-ratio pitch-shifting, we evaluate these metrics using ratios of .71, 1 (unmodified), and 1.41.

We perform objective evaluation of variable-ratio pitch-shifting on RAVDESS. RAVDESS contains an English speech dataset with 24 speakers saying two sentences with many different, expressive prosodies. We select pairs of utterances where the same speaker says the same sentence with different pitch and phoneme durations. We use pitch-shifting and time-stretching to make one utterance in a pair have the pitch and phoneme durations of the other. We use the pitch of the target utterance as ground truth for evaluation. We create 5 pairs each from 20 speakers, for a total of 199 pairs (one speaker only produced 4 pairs) from 277 unique utterances. Given the multimodality of English prosody, this is a suitable prosody transfer task, producing pitch-shifting ratios between .4 and 2.5 and time-stretching ratios between .25 and 4.

We perform objective evaluation of constant-ratio pitch-shifting on VCTK, DAPS, and RAVDESS. For VCTK, we use four utterances from eight seen speakers and four utterances from eight unseen speakers. For DAPS, we use ten utterances from ten speakers. For RAVDESS, we use 100 utterances randomly selected among the 277 used for variable-ratio evaluation.

We perform three ablations to evaluate our methods proposed in sections \ref{sec:pitch}-\ref{sec:sample}. For section \ref{sec:pitch}, we use the YIN pitch and non-uniform bin spacing of the original LPCNet. We set all pitch values less than 63 Hz to 63 Hz, as this representation cannot represent frequencies below this point. We use YIN to evaluate the pitch accuracy of this ablation. To ablate \ref{sec:augment}, we remove our proposed resampling augmentation. For section \ref{sec:sample}, we remove the distribution threshold (i.e., we set $T = 0$) to demonstrate the importance of this parameter.

\begin{figure*}[t]
    \centering
    \includegraphics[width=\textwidth]{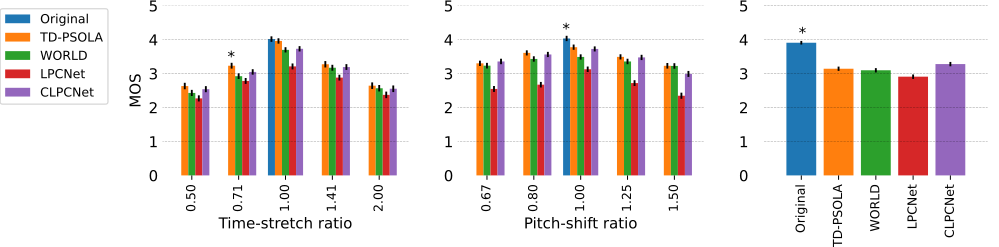}
    \caption{Subjective mean opinion scores (MOS) for (\textbf{left}) constant-ratio time-stretching, (\textbf{center}) constant-ratio pitch-shifting, and (\textbf{right}) variable-ratio pitch-shifting and time-stretching. Asterisks indicate statistically significant winners of two-sided \textit{t}-tests with $p=0.05$.}
    \label{fig:subjective}
    \vspace{-3mm}
\end{figure*}

\subsection{Subjective evaluation}
\label{sec:subje}

We report the results of subjective experiments designed to evaluate the ability of CLPCNet to perform both constant- and variable-ratio pitch-shifting and time-stretching. All experiments are mean opinion score (MOS) tests conducted on Amazon Mechanical Turk with a scale from 1 (worst) to 5 (best).
For all experiments, we test the audio quality using five conditions: (1) the original audio, (2) TD-PSOLA, (3), WORLD, (4) LPCNet, and (5) CLPCNet. We perform variable-ratio evaluation on the RAVDESS dataset and constant-ratio evaluation on DAPS, using the same examples as in our objective evaluation. We evaluate pitch-shifting at constant ratios of .67, .80, 1, 1.25, and 1.5. We evaluate time-stretching at constant ratios of .50, .71, 1, 1.41, and 2.

\begin{table}[t]
\centering
\begin{tabular}{ll|crc|}
\textbf{Method} & \textbf{Rate} & \textbf{F1$\uparrow$} & \textbf{RMS$\downarrow$} & \textbf{GPE$\downarrow$}\\
\hline
\hline
      & 0.71 & .995 & 28.9 & .029 \\
TD-PSOLA & 1.00 & .999 & 8.4 & .000 \\
      & 1.41 & .996 & 19.8 & .026 \\
\rowcolor{gray!12}
      & 0.71 & .935 & 16.1 & .017 \\
\rowcolor{gray!12}
WORLD & 1.00 & .935 & 19.2 & .027 \\
\rowcolor{gray!12}
      & 1.41 & .935 & 16.2 & .016 \\
       & 0.71 & .779 & 434.0 & .290 \\
LPCNet & 1.00 & .791 & 79.5 & .101 \\
       & 1.41 & .786 & 163.0 & .239 \\
\rowcolor{gray!12}
        & 0.71 & .942 & 66.4 & .228 \\
\rowcolor{gray!12}
CLPCNet & 1.00 & .945 & 21.6 & .040 \\
\rowcolor{gray!12}
        & 1.41 & .941 & 119.0 & .304 \\
            & 0.71 & .914 & 250.0 & .389 \\
\quad - pitch (\ref{sec:pitch}) & 1.00 & .923 & 75.3 & .140 \\
            & 1.41 & .920 & 221.0 & .341 \\
\rowcolor{gray!12}
            & 0.71 & .931 & 209.0 & .332 \\
\rowcolor{gray!12}
\quad - augmentation (\ref{sec:augment}) & 1.00 & .938 & 25.0 & .049 \\
\rowcolor{gray!12}
            & 1.41 & .933 & 177.0 & .356 \\
            & 0.71 & .596 & 44.9 & .114 \\
\quad - sampling (\ref{sec:sample}) & 1.00 & .596 & 17.1 & .028 \\
            & 1.41 & .594 & 84.1 & .152 \\
\end{tabular}
\caption{Objective evaluation of pitch-shifting with baselines and three ablations on DAPS for three constant ratios.}
\label{tab:pitch}
\vspace{-4mm}
\end{table}

\begin{table}
\centering
\begin{tabular}{ll|crc|}
\textbf{Dataset} & \textbf{Rate} & \textbf{F1$\uparrow$} & \textbf{RMS$\downarrow$} & \textbf{GPE$\downarrow$} \\
\hline
\hline
            & 0.71 & .931 & 49.5 & .133 \\
VCTK (seen) & 1.00 & .931 & 20.0 & .035 \\
            & 1.41 & .925 & 81.1 & .185 \\
\rowcolor{gray!12}
              & 0.71 & .924 & 47.0 & .117 \\
\rowcolor{gray!12}
VCTK (unseen) & 1.00 & .925 & 17.6 & .024 \\
\rowcolor{gray!12}
              & 1.41 & .922 & 85.7 & .160 \\
        & 0.71 & .943 & 58.2 & .128 \\
RAVDESS & 1.00 & .945 & 17.6 & .029 \\
        & 1.41 & .942 & 104.0 & .199 \\
\end{tabular}
\caption{Objective evaluation of constant-ratio pitch-shifting with CLPCNet on VCTK and RAVDESS for three constant ratios.}
\label{tab:gener}
\end{table}

\begin{table}[t]
\centering
\begin{tabular}{lccc}
\textbf{Method} & \textbf{F1$\uparrow$} & \textbf{RMS$\downarrow$} & \textbf{GPE$\downarrow$} \\
\hline
\hline
TD-PSOLA  & .901 & 14.2 & .015 \\
\rowcolor{gray!12}
WORLD & .893 & 14.0 & .016 \\
LPCNet & .748 & 99.3 & .104 \\
\rowcolor{gray!12}
CLPCNet & .883 & 58.3 & .149 \\
\end{tabular}
\caption{Objective evaluation of variable-ratio pitch-shifting on the RAVDESS dataset.}
\label{tab:var}
\vspace{-4mm}
\end{table}

\section{Results}

In our objective evaluation, we find that CLPCNet significantly improves the F1 and RMS of pitch-shifting (see Tables~\ref{tab:pitch} and~\ref{tab:var}) compared to the original LPCNet. The pitch-shifting accuracy is less than TD-PSOLA or WORLD. However, prior studies have shown that humans do not register variations in pitch less than 150 cents as a distinct prosody~\cite{rietveld1985relation}, making CLPCNet suitable for prosody editing. Our ablations highlight crucial design decisions for high-quality pitch-shifting, but do not completely explain the performance gap between CLPCNet and LPCNet. The remaining gap is due to increasing the amount of training data, removing the original data augmentation, and removing the sparsity constraint on the GRU.

While more direct comparison is needed, our F1 on clean speech data substantially outperforms those reported by previous state-of-the-art neural methods such as PS-WaveNet~\cite{morrison2020controllable}, QP-PWG~\cite{Wu_2021}, and uSFGAN~\cite{yoneyama2021unified}, and our RMS compares favorably or better. \textbf{Note that a higher F1 score makes having a low RMS more difficult, as more voiced frames are being evaluated (e.g., compare CLPCNet with and without the sampling ablation in Table~\ref{tab:pitch}).}

We analyze the results of our subjective evaluation (Figure~\ref{fig:subjective}) using two-sided $t$-tests with a $p$-value of 0.05. We find that CLPCNet outperforms LPCNet on all conditions. CLPCNet outperforms WORLD on constant-ratio time-stretching for ratios less than one, as well as for constant-ratio pitch-shifting for ratios less than 1.5. WORLD outperforms CLPCNet only for pitch-shifting with a ratio of 1.5. TD-PSOLA outperforms CLPCNet for time-stretching with a ratio of 0.71 and pitch-shifting with a ratio of 1.5. However, TD-PSOLA is non-parametric, and cannot be used for, e.g., speech coding or vocoding. CLPCNet outperforms all conditions on variable-ratio pitch-shifting and time-stretching (e.g., prosody editing), but does not match the quality of the original recording.


\section{Conclusion}

Modern speech editing software necessitates high-quality, natural-sounding speech manipulation. In this paper, we introduce CLPCNet, an improved LPCNet vocoder that makes significant progress towards this goal. In objective evaluation, we show that CLPCNet exhibits pitch-shifting accuracy suitable for speech prosody editing. In subjective evaluation, we show that the quality of pitch-shifting and time-stretching with CLPCNet is comparable or better than LPCNet, TD-PSOLA, and WORLD.


\vfill\pagebreak

\bibliographystyle{IEEEbib}
\bibliography{refs}

\end{document}